\documentclass[manuscript]{aastex63}

\newcommand{\sm}{$M_\odot$}

\newcommand{\tbol}{$T_{\rm bol}$}

\newcommand{\iras}{IRAS 15398$-$3359}
\newcommand{\hhco}{H$_{2}$CO}

\newcommand{\co}{C$^{18}$O}
\newcommand{\ccchh}{c-C$_3$H$_2$}

\newcommand{\meta}{CH$_3$OH}

\newcommand{\hthcop}{H$^{13}$CO$^+$}

\newcommand{\kms}{km s$^{-1}$}
\newcommand{\mjybeam}{mJy beam$^{-1}$}

\revised{\today}
\submitjournal{ApJ}

\shorttitle{}
\shortauthors{okoda et al.}

\begin{document}

\title{FAUST II. Discovery of a Secondary Outflow in \iras: Variability in Outflow Direction during the Earliest Stage of Star Formation?}
\footnote{}

\correspondingauthor{Yuki Okoda}
\email{okoda@taurus.phys.s.u-tokyo.ac.jp}

\author{Yuki Okoda}
\affiliation{Department of Physics, The University of Tokyo, 7-3-1, Hongo, Bunkyo-ku, Tokyo 113-0033, Japan}

\author{Yoko Oya}
\affiliation{Department of Physics, The University of Tokyo, 7-3-1, Hongo, Bunkyo-ku, Tokyo 113-0033, Japan}
\affiliation{Research Center for the Early Universe, The University of Tokyo, 7-3-1, Hongo, Bunkyo-ku, Tokyo 113-0033, Japan}

\author{Logan Francis}
\affiliation{NRC Herzberg Astronomy and Astrophysics, 5071 West Saanich Road, Victoria, BC, V9E 2E7, Canada}
 \affiliation{Department of Physics and Astronomy, University of Victoria, Victoria, BC, V8P 5C2, Canada}

\author{Doug Johnstone}
\affiliation{NRC Herzberg Astronomy and Astrophysics, 5071 West Saanich Road, Victoria, BC, V9E 2E7, Canada}
\affiliation{Department of Physics and Astronomy, University of Victoria, Victoria, BC, V8P 5C2, Canada}

\author{Shu-ichiro Inutsuka}
\affiliation{Department of Physics, Nagoya University, Furo-cho, Chikusa-ku, Nagoya, Aichi 464-8602, Japan}

\author{Cecilia Ceccarelli}
\affiliation{Univ. Grenoble Alpes, CNRS, IPAG, 38000 Grenoble, France}
\author{Claudio Codella}
\affiliation{Univ. Grenoble Alpes, CNRS, IPAG, 38000 Grenoble, France}
\affiliation{INAF, Osservatorio Astrofisico di Arcetri, Largo E. Fermi 5, I-50125, Firenze, Italy}
\author{Claire Chandler}
\affiliation{National Radio Astronomy Observatory, PO Box O, Socorro, NM 87801, USA}
\author{Nami Sakai}
\affiliation{RIKEN Cluster for Pioneering Research, 2-1, Hirosawa, Wako-shi, Saitama 351-0198, Japan}
\author{Yuri Aikawa}
\affiliation{Department of Astronomy, The University of Tokyo, 7-3-1 Hongo, Bunkyo-ku, Tokyo 113-0033, Japan}
\author{Felipe Alves}
\affiliation{Center for Astrochemical Studies, Max-Planck-Institut f\"{u}r extraterrestrische Physik (MPE), Gie$\beta$enbachstr. 1, D-85741 Garching, Germany}
\author{Nadia Balucani}
\affiliation{Department of Chemistry, Biology, and Biotechnology, The University of Perugia, Via Elce di Sotto 8, 06123 Perugia, Italy}
\author{Eleonora Bianchi}
\affiliation{Univ. Grenoble Alpes, CNRS, IPAG, 38000 Grenoble, France}
\author{Mathilde Bouvier}
\affiliation{Univ. Grenoble Alpes, CNRS, IPAG, 38000 Grenoble, France}
\author{Paola Caselli}
\affiliation{Center for Astrochemical Studies, Max-Planck-Institut f\"{u}r extraterrestrische Physik (MPE), Gie$\beta$enbachstr. 1, D-85741 Garching, Germany}
\author{Emmanuel Caux}
\affiliation{IRAP, Universit\'{e} de Toulouse, CNRS, CNES, UPS, Toulouse, France}
\author{Steven Charnley}
\affiliation{Astrochemistry Laboratory, Code 691, NASA Goddard Space Flight Center, 8800 Greenbelt Road, Greenbelt, MD 20771, USA}
\author{Spandan Choudhury}
\affiliation{Center for Astrochemical Studies, Max-Planck-Institut f\"{u}r extraterrestrische Physik (MPE), Gie$\beta$enbachstr. 1, D-85741 Garching, Germany}
\author{Marta De Simone}
\affiliation{Univ. Grenoble Alpes, CNRS, IPAG, 38000 Grenoble, France}
\author{Francois Dulieu}
\affiliation{CY Cergy Paris Universit\'{e}, Sorbonne Universit\'{e}, Observatoire de Paris, PSL University, CNRS, LERMA, F-95000, Cergy, France}
\author{Aurora Dur\'{a}n}
\affiliation{Instituto de Radioastronom\'{i}a y Astrof\'{i}sica , Universidad Nacional Aut\'{o}noma de M\'{e}xico, A.P. 3-72 (Xangari), 8701, Morelia, Mexico}
\author{Lucy Evans}
\affiliation{IRAP, Universit\'{e} de Toulouse, CNRS, CNES, UPS, Toulouse, France}
\affiliation{INAF, Osservatorio Astrofisico di Arcetri, Largo E. Fermi 5, I-50125, Firenze, Italy}
\author{C\'{e}cile Favre}
\affiliation{Univ. Grenoble Alpes, CNRS, IPAG, 38000 Grenoble, France}
\author{Davide Fedele}
\affiliation{INAF, Osservatorio Astrofisico di Arcetri, Largo E. Fermi 5, I-50125, Firenze, Italy}
\author{Siyi Feng}
\affiliation{CAS Key Laboratory of FAST, National Astronomical Observatory of China, Datun Road 20, Chaoyang, Beijing, 100012, P. R. China}
\affiliation{National Astronomical Observatory of Japan, National Institutes of Natural Sciences, 2-21-1 Osawa, Mitaka, Tokyo 181-8588, Japan}
\affiliation{Academia Sinica Institute of Astronomy and Astrophysics, No.1, Sec. 4, Roosevelt Rd, Taipei 10617, Taiwan, Republic of China}
\author{Francesco Fontani}
\affiliation{INAF, Osservatorio Astrofisico di Arcetri, Largo E. Fermi 5, I-50125, Firenze, Italy}
\affiliation{Center for Astrochemical Studies, Max-Planck-Institut f\"{u}r extraterrestrische Physik (MPE), Gie$\beta$enbachstr. 1, D-85741 Garching, Germany}
\author{Tetsuya Hama}
\affiliation{Komaba Institute for Science, The University of Tokyo, 3-8-1 Komaba, Meguro, Tokyo 153-8902, Japan}
\affiliation{Department of Basic Science, The University of Tokyo, 3-8-1 Komaba, Meguro, Tokyo 153-8902, Japan}
\author{Tomoyuki Hanawa}
\affiliation{Center for Frontier Science, Chiba University, 1-33 Yayoi-cho, Inage-ku, Chiba 263-8522, Japan}
\author{Eric Herbst}
\affiliation{Department of Chemistry, University of Virginia, McCormick Road, PO Box 400319, Charlottesville, VA 22904, USA}
\author{Tomoya Hirota}
\affiliation{National Astronomical Observatory of Japan, Osawa 2-21-1, Mitaka-shi, Tokyo 181-8588, Japan}
\author{Muneaki Imai}
\affiliation{Department of Physics, The University of Tokyo, 7-3-1, Hongo, Bunkyo-ku, Tokyo 113-0033, Japan}
\author{Andrea Isella}
\affiliation{Department of Physics and Astronomy, Rice University, 6100 Main Street, MS-108, Houston, TX 77005, USA}
\author{Izaskun J\'{i}menez-Serra}
\affiliation{Centro de Astrobiolog\'{\i}a (CSIC-INTA), Ctra. de Torrej\'on a Ajalvir, km 4, 28850, Torrej\'on de Ardoz, Spain}
\author{Claudine Kahane}
\affiliation{Univ. Grenoble Alpes, CNRS, IPAG, 38000 Grenoble, France}
\author{Bertrand Lefloch}
\affiliation{Univ. Grenoble Alpes, CNRS, IPAG, 38000 Grenoble, France}
\author{Laurent Loinard}
\affiliation{Instituto de Radioastronom\'{i}a y Astrof\'{i}sica , Universidad Nacional Aut\'{o}noma de M\'{e}xico, A.P. 3-72 (Xangari), 8701, Morelia, Mexico}
\affiliation{Instituto de Astronom\'{i}a, Universidad Nacional Aut\'{o}noma de M\'{e}xico, Ciudad Universitaria, A.P. 70-264, Cuidad de M\'{e}xico 04510, Mexico}
\author{Ana L\'{o}pez-Sepulcre}
\affiliation{Univ. Grenoble Alpes, CNRS, IPAG, 38000 Grenoble, France}
\affiliation{Institut de Radioastronomie Millim\'{e}trique, 38406 Saint-Martin d'H$\grave{e}$res, France}
\author{Luke T. Maud}
\affiliation{European Southern Observatory, Karl-Schwarzschild Str. 2, 85748 Garching bei M\"{u}nchen, Germany}
\author{Maria Jose Maureira}
\affiliation{Center for Astrochemical Studies, Max-Planck-Institut f\"{u}r extraterrestrische Physik (MPE), Gie$\beta$enbachstr. 1, D-85741 Garching, Germany}
\author{Francois Menard}
\affiliation{Univ. Grenoble Alpes, CNRS, IPAG, 38000 Grenoble, France}
\author{Seyma Mercimek}
\affiliation{INAF, Osservatorio Astrofisico di Arcetri, Largo E. Fermi 5, I-50125, Firenze, Italy}
\affiliation{Universit$\grave{a}$ degli Studi di Firenze, Dipartimento di Fisica e Astronomia, via G. Sansone 1, 50019 Sesto Fiorentino, Italy}
\author{Anna Miotello}
\affiliation{European Southern Observatory, Karl-Schwarzschild Str. 2, 85748 Garching bei M\"{u}nchen, Germany}
\author{George Moellenbrock}
\affiliation{National Radio Astronomy Observatory, PO Box O, Socorro, NM 87801, USA}
\author{Shoji Mori}
\affiliation{Department of Astronomy, The University of Tokyo, 7-3-1, Hongo, Bunkyo-ku, Tokyo 113-0033, Japan}
\author{Nadia M. Murillo}
\affiliation{RIKEN Cluster for Pioneering Research, 2-1, Hirosawa, Wako-shi, Saitama 351-0198, Japan}
\author{Riouhei Nakatani}
\affiliation{RIKEN Cluster for Pioneering Research, 2-1, Hirosawa, Wako-shi, Saitama 351-0198, Japan}
\author{Hideko Nomura}
\affiliation{Division of Science, National Astronomical Observatory of Japan, 2-21-1 Osawa, Mitaka, Tokyo 181-8588, Japan}
\author{Yasuhiro Oba}
\affiliation{Institute of Low Temperature Science, Hokkaido University, N19W8, Kita-ku, Sapporo, Hokkaido 060-0819, Japan}
\author{Ross O'Donoghue}
\affiliation{Department of Physics and Astronomy, University College London, Gower Street, London, WC1E 6BT, UK}
\author{Satoshi Ohashi}
\affiliation{RIKEN Cluster for Pioneering Research, 2-1, Hirosawa, Wako-shi, Saitama 351-0198, Japan}
\author{Juan Ospina-Zamudio}
\affiliation{Univ. Grenoble Alpes, CNRS, IPAG, 38000 Grenoble, France}
\author{Jaime Pineda}
\affiliation{Center for Astrochemical Studies, Max-Planck-Institut f\"{u}r extraterrestrische Physik (MPE), Gie$\beta$enbachstr. 1, D-85741 Garching, Germany}
\author{Linda Podio}
\affiliation{INAF, Osservatorio Astrofisico di Arcetri, Largo E. Fermi 5, I-50125, Firenze, Italy}
\author{Albert Rimola}
\affiliation{Departament de Qu\'{i}mica, Universitat Aut$\grave{o}$noma de Barcelona, 08193 Bellaterra, Spain}
\author{Takeshi Sakai}
\affiliation{Graduate School of Informatics and Engineering, The University of Electro-Communications, Chofu, Tokyo 182-8585, Japan}
\author{Dominique Segura Cox}
\affiliation{Center for Astrochemical Studies, Max-Planck-Institut f\"{u}r extraterrestrische Physik (MPE), Gie$\beta$enbachstr. 1, D-85741 Garching, Germany}
\author{Yancy Shirley}
\affiliation{Steward Observatory, 933 N Cherry Ave., Tucson, AZ 85721 USA}
\author{Brian Svoboda}
\affiliation{National Radio Astronomy Observatory, PO Box O, Socorro, NM 87801, USA}
\affiliation{Jansky Fellow of the National Radio Astronomy Observatory.}
\author{Vianney Taquet}
\affiliation{INAF, Osservatorio Astrofisico di Arcetri, Largo E. Fermi 5, I-50125, Firenze, Italy}
\author{Leonardo Testi}
\affiliation{European Southern Observatory, Karl-Schwarzschild Str. 2, 85748 Garching bei M\"{u}nchen, Germany}
\affiliation{INAF, Osservatorio Astrofisico di Arcetri, Largo E. Fermi 5, I-50125, Firenze, Italy}
\author{Charlotte Vastel}
\affiliation{IRAP, Universit\'{e} de Toulouse, CNRS, CNES, UPS, Toulouse, France}
\author{Serena Viti}
\affiliation{Department of Physics and Astronomy, University College London, Gower Street, London, WC1E 6BT, UK}
\author{Naoki Watanabe}
\affiliation{Institute of Low Temperature Science, Hokkaido University, N19W8, Kita-ku, Sapporo, Hokkaido 060-0819, Japan}
\author{Yoshimasa Watanabe}
\affiliation{Materials Science and Engineering, College of Engineering, Shibaura Institute of Technology, 3-7-5 Toyosu, Koto-ku, Tokyo 135-8548, Japan}
\author{Arezu Witzel}
\affiliation{Univ. Grenoble Alpes, CNRS, IPAG, 38000 Grenoble, France}
\author{Ci Xue}
\affiliation{Department of Chemistry, University of Virginia, McCormick Road, PO Box 400319, Charlottesville, VA 22904, USA}
\author{Yichen Zhang}
\affiliation{RIKEN Cluster for Pioneering Research, 2-1, Hirosawa, Wako-shi, Saitama 351-0198, Japan}
\author{Bo Zhao}
\affiliation{Center for Astrochemical Studies, Max-Planck-Institut f\"{u}r extraterrestrische Physik (MPE), Gie$\beta$enbachstr. 1, D-85741 Garching, Germany}
\author{Satoshi Yamamoto}
\affiliation{Department of Physics, The University of Tokyo, 7-3-1, Hongo, Bunkyo-ku, Tokyo 113-0033, Japan}
\affiliation{Research Center for the Early Universe, The University of Tokyo, 7-3-1, Hongo, Bunkyo-ku, Tokyo 113-0033, Japan}

\begin{abstract}

\par 
We have observed the very low-mass Class 0 protostar \iras\ at scales ranging from 50 au to 1800 au, as part of the ALMA Large Program FAUST.
We uncover a linear feature, visible in \hhco, SO, and \co\ line emission, which extends from the source along a direction almost perpendicular to the known active outflow.
Molecular line emission from \hhco, SO, SiO,  and CH$_3$OH further reveals an arc-like structure connected to the outer end of the linear feature and separated from the protostar, \iras, by 1200 au.
The arc-like structure is blue-shifted with respect to the systemic velocity.
A velocity gradient of 1.2 \kms\ over 1200 au along the linear feature seen in the \hhco\ emission connects the protostar and the arc-like structure kinematically. 
SO, SiO, and \meta\ are known to trace shocks, and we interpret the arc-like structure as a relic shock region produced by an outflow previously launched by \iras.
The velocity gradient along the linear structure can be explained as relic outflow motion.
The origins of the newly observed arc-like structure and extended linear feature are discussed in relation to turbulent motions within the protostellar core and episodic accretion events during the earliest stage of protostellar evolution.

\end{abstract}

\keywords{ISM: individual objects (IRAS 15398$-$3359) - ISM: molecules - stars: formation}


\section{Introduction}
Millimeter/submillimeter-wave (mm/sub-mm) observations are revealing the complex physical and chemical nature of low-mass protostellar systems during their earliest evolutionary stage. 
For instance, protostellar accretion bursts resulting in a large instantaneous increase of the protostellar luminosity and subsequent heating of the protostellar envelope have been suggested for Class 0 sources based on the observed spatial distribution of molecular species \citep{Jorgensen et al.(2013), Hsieh et al.(2018), Hsieh et al.(2019)}. 
Furthermore, complex structure, consisting of arc-like features and dense clumps, has been reported around the very young protostellar core, L1521F \citep{Tokuda et al.(2014), Favre et al.(2020)}, and for the first hydrostatic core candidate, Chamaeleon-MMS1 \citep{Busch et al.(2020)}.
Because the protostars are deeply embedded in their parent cores, interactions between a protostellar outflow and surrounding gas may contribute to such complicated morphologies.
Thus, detailing these structures will provide us with an important clue to elucidating a dynamic feature in the earliest stage of protostellar evolution.
Given these circumstances, it is increasingly important to explore carefully and in detail the earliest stage of star formation for specific sources. 

\par \iras\ is a young low-mass Class 0 protostellar source \citep*[\tbol$=$44 K;][]{Jorgensen et al.(2013)} located in the Lupus 1 molecular cloud \citep*[$d=$156 pc;][]{Dzib et al.(2018)}.
A molecular outflow driven from the protostar has been found via single-dish observations in CO emission \citep{Tachihara et al.(1996), van Kempen et al.(2009)}. 
A variety of unsaturated carbon-chain molecules such as CCH, $\rm C_{4}$H, and C$\rm H_3$CCH are abundant on scales of a few thousand au around the protostar. Hence, this source is classified as a warm carbon-chain chemistry (WCCC) source \citep{Sakai et al.(2009), Sakai & Yamamoto(2013)}.

\par High resolution observations with the Atacama Large Millimeter/Submillimeter Array (ALMA) toward this source have revealed the outflow structure and the disk/envelope system.
A bipolar outflow extending from the northeast to the southwest (P.A. 220\degr) was reported in CCH and \hhco\ line emission by \cite{Oya et al.(2014)}.
The authors evaluated the inclination angle of the outflow axis to be 70\degr\ (0\degr\ for a pole-on configuration) based on their outflow model.
Thus, the outflow is launched almost parallel to the plane of the sky.
Submillimeter Array (SMA) CO emission observations confirm this outflow feature \citep{Bjerkeli et al.(2016a)}.
The outflow dynamical timescale was derived to be very short, 10$^{2}-10^{3}$ yr \citep{Oya et al.(2014), Yildiz et al.(2015), Bjerkeli et al.(2016a)}.
Significantly, this protostar has a very low dynamical mass.
\cite{Oya et al.(2014)} and \cite{Yen et al.(2017)} derived the upper limit of the protostellar mass to be 0.09 \sm\ and 0.01 \sm\ using the observed velocity structure of the \hhco\ and \co\ line emission, respectively, at a resolution of 0\farcs5 ($\sim$80 au).
More recently, a disk structure with Keplerian rotation was observed in SO line emission at a higher angular resolution (0\farcs2; $\sim$30 au) and the protostellar mass was re-evaluated to be 0.007$^{+0.004}_{-0.003}$ \sm\ \citep{Okoda et al.(2018)}.
The authors reported that the central velocity of the disk structure, 5.5 \kms, is slightly shifted from the systemic velocity of the protostellar core, $\sim$5.2 \kms\ \citep{Oya et al.(2014), Yen et al.(2017)}.
They also estimated the disk mass to be between 0.006 \sm\ and 0.001 \sm\ from 1.2 mm dust continuum emission, assuming dust
temperature of 20 K and 100 K, respectively.
Despite the measurement uncertainties, the disk thus appears to have a lower mass than the mass of the protostar.
\cite{Kristensen et al.(2012)} and \cite{Jorgensen et al.(2013)} previously reported that the envelope mass is 0.5$-$1.2 \sm. 
Therefore, \iras\ should be in the earliest stage of protostellar evolution.

\par Due to these unique physical and chemical characteristics, \iras\ was selected as a target of the ALMA Large Program FAUST (Fifty AU STudy of the chemistry in the disk/envelope system of Solar-like protostars\footnote{\url{http://faust-alma.riken.jp}}).
This program aims at revealing the physical and chemical structure of thirteen nearby protostars ($d=$137$-$235 pc), at scales from a few 1000 au down to 50 au, by observing various molecular lines.
In addition to lines of fundamental molecules (e.g., \co\ and \hhco) and complex organic molecules, some shock tracer molecules, such as SO, SiO, and \meta, are included in this program to reveal possible interactions between the outflow and the ambient gas: these species are thought to be liberated from dust grains or to be produced in the gas phase within shocked regions.

\par Taking advantage of the chemical diagnostic power of FAUST obtained by observing these shock tracer lines with high sensitivity, we have found a secondary outflow feature launched from \iras, which provides a novel insight into the earliest stage of protostellar evolution. 
In Section \ref{sec-obs} we present the relevant observations. Next, in Section \ref{sec-res} we discuss the results of our spatial and spectral investigation. We analyze and discuss these findings in Sections \ref{ana} and \ref{sec-disc}, respectively, before concluding the paper in Section \ref{sec-conc}.

\section{Observations}
\label{sec-obs}
Single field observations for \iras\ were conducted between October 2018 and January 2019 as part of the ALMA Large Program FAUST.
The parameters of observations are summarized in Table \ref{observations}.
The molecular lines analyzed in the frequency range from 217 GHz to 220 GHz (Band 6) are listed in Table \ref{line}.
We used the 12-m array data from the two different configurations (C43-5 and C43-2 for sparse and compact configurations, respectively) and the 7-m array data of Atacama Compact Array (ACA/Morita Array), combining these visisbilty data in the UV plane.
In total, the baseline lengths range from 7.43 m to 1310.74 m.
The adopted field center was taken to be  ($\alpha_{2000}$, $\delta_{2000}$)= (15\fh43\fm02\fs242, $-$34\arcdeg 09\arcmin 06\farcs805), which is close to the protostellar position.
The backend correlator for the molecular line observations was set to a resolution of 122 kHz and a bandwidth of 62.5 MHz.
The data were reduced in Common Astronomy Software Applications package (CASA) 5.4.1 \citep{McMullin et al.(2007)} using a modified version of the ALMA calibration pipeline and an additional in-house calibration routine (Mollenbrock et al. in prep) to correct for the $T_{\rm sys}$ and spectral line data normalization\footnote{\url{https://help.almascience.org/index.php?/Knowledgebase/Article/View/419}}. 
Self-calibration was carried out using line-free continuum emission, for each configuration. The complex gain corrections derived from the self calibration were then applied to all channels in the data, and the continuum model derived from the self calibration was subtracted from the data to produce continuum-subtracted line data. A self-calibration technique was also used to align both amplitudes and phases (i.e., positions) across the multiple configurations.
Images were prepared by using the {\it tclean} task in CASA, where Briggs weighting with a robustness parameter of 0.5 was employed. 
The primary beam correction was applied to all the images presented in this paper.
Since the maximum recoverable scales are 12\farcs8, any structures extended more than that size could be resolved-out.
The root mean square (rms) noise levels for \hhco, SiO, \meta, and \co\ and for SO are 2 mJy beam$^{-1}$channel$^{-1}$ and  3 mJy beam$^{-1}$channel$^{-1}$, respectively. 
The original synthesized beam sizes are summarized in Table \ref{line}.
The uncertainty in the absolute flux density scale is estimated to be 10 \%\ \citep{Francis et al.(2020)}.

\section{A Possible Secondary Outflow}
\label{sec-res}
\subsection{Primary Outflow along the Northeast to Southwest Direction}
\par Figures \ref{moment_1}a and \ref{moment_2}a show the moment 0 maps of the \hhco\ and SO line emission, respectively.
The outflow structure along the northeast to southwest axis (P.A. 220\degr) is seen in the \hhco\ emission, as reported previously \citep{Oya et al.(2014), Bjerkeli et al.(2016a)}.
Contrastingly, the SO emission is concentrated around the protostar \citep{Okoda et al.(2018), Okoda et al.(2020)} with little emission within the outflow except for a localized knot \citep*[Blob D in][]{Okoda et al.(2020)} seen in the southwestern lobe, which could be formed by an impact of the outflow on ambient gas.
The \meta\ and \co\ line emission trace part of the outflow in Figures \ref{moment_2}b and \ref{moment_2}c.
The knot can also be seen in the \meta\ emission \citep*[See also][]{Okoda et al.(2020)}.
Along with these previously known structures, we have found an additional spatial feature extending toward the southeast and northwest of \iras, which was not reported in previous studies.

\subsection{Arc-Like Structure toward Southeast Direction}
\par Part of the \hhco\ line emission and a majority of the SO line emission are extended along the southeastern direction (P.A. 140\degr), which is also close to the disk/envelope direction (P.A. 130\degr) of the \iras\ protostar reported by \cite{Oya et al.(2014)}, \cite{Yen et al.(2017)}, and \cite{Okoda et al.(2018)}.
The \hhco\ emission appears to bend toward the south about 8$''$ ($\sim$1200 au) from the protostar. 
The moment 0 map of the SO emission shows an arc-like structure around the southeastern part (Figure \ref{moment_2}a), where the northern tip of the arc corresponds to the bending point seen in the \hhco\ emission.
Although the arc-like structure is near the edge of the field of view for the ALMA 12 m data, it lies within the field of view of ACA data.
Hence, the observed structure is real.

\par Along with the \hhco\ and SO line emission, the SiO and \meta\ line emission is found to trace the arc-like structure around the southeastern part.
In contrast to the \hhco\ and SO line emission, the SiO emission is not detected toward the protostar, and mainly traces the southeastern part of the arc-like structure (Figure \ref{moment_2}d).
These features are also apparent in the molecular line profiles (Figure \ref{spectral}) toward four positions within the arc-like structure and at the protostellar position, as indicated in Figure \ref{moment_2}. 
The  line width of the SiO emission is slightly broader at position D (0.9 \kms) than at positions A, B, and C (0.3$-$0.8 \kms).
Faint emission from \meta\ is also detected in part of the arc-like structure in addition to the previously observed outflow cavity wall (Figure \ref{moment_2}b).
While the \meta\ emission looks faint in the moment 0 map, its spectrum is clearly detected in the arc-like structure, as revealed in Figure \ref{spectral}.

\par In Figure \ref{channel_1}, we present the velocity channel maps of the \hhco, SO, SiO, and \meta\ line emission from the 3.9 \kms\ panel to the 5.5 \kms\ panel, where the systemic velocity of the protostellar core is 5.2 \kms\ \citep{Yen et al.(2017)}.
The velocity channel maps reveal the arc-like structure.
The \meta\ emission is seen clearly in the velocity maps at 4.7 \kms\ and 5.1 \kms. 
On the other hand, the arc-like structure is not evident in the \co\ emission, as shown in the line profiles and the velocity channel maps (Figures \ref{spectral} and \ref{channel_1} e).
The \co\ emission usually traces cold and dense clumps of gas with relatively high column density.
Although marginal emission may be detected around the southeastern extent in the map of 5.5 \kms, the low \co\ column density (calculated in Section 4.2)
ensures that the arc-like structure is not part of an adjacent prestellar core.

\subsection{Shocks in the Arc-Like Structure}
\par The SO, SiO, and \meta\ line emission are known to be enhanced in the gas phase by shocks and are often detected in active outflow-shocked regions \citep*[e.g.,][]{Bachiller(1997)}, although SO and \meta\ are also present in quiescent starless cores \citep*[e.g.,][]{Spezzano et al.(2017), Punanova et al.(2018)}.
A typical example of such a shock region in a  low-mass star-forming region occurs in the outflow driven from the Class 0 protostar, IRAS 20386+6751, in L1157 \citep{Mikami et al.(1992), Gueth et al.(1996), Bachiller(1997)}.
Outflows often produce shocked emission at locations where they interact with ambient gas.
In these outflow-shocked regions, the chemical composition of the gas is drastically changed, mainly due to liberation of molecules from dust grains and subsequent gas-phase reactions.
In L1157, the shocked region in the blue-shifted outflow lobe (L1157-B1) is traced by SO, SiO, \meta, \hhco, HCN, CN, and SO$_2$ line emission \citep{Mikami et al.(1992), Gueth et al.(1996), Bachiller(1997), Podio et al.(2017), Codella et al.(2010), Codella et al.(2020), Feng et al.(2020)}.
Enhancement of the same molecules has also been reported for other protostellar outflows: {\bf e.g., L1448-C (L1448-mm); \cite{Jimenez-Serra et al.(2005), Hirano et al.(2010)}, HH211; \cite{Hirano et al.(2006)}, NGCC1333-IRAS4A; \cite{Wakelam et al.(2005)}, BHR71; \cite{Gusdorf et al.(2015)}.}
Since SiO is a robust shock tracer, the observed {\bf arc}-like structure to the southeast of \iras\ should also be a shocked region.
We also detect enhancement in the abundance of other species, SO and \meta\ (Figures \ref{moment_2}a and 2b), which is consistent with the enhancement reported toward the shocked region, L1157-B1.
Furthermore, the morphology resembles a bow shock, somewhat similar to those revealed by the simulations \citep{Smith et al.(1997), Lee et al.(2001)} and the observation \citep*[HH46/47;][]{Arce et al.(2013)}.
\par The line width of the SiO emission in typical shocked regions is generally broad \cite*[$\sim$10 \kms; e.g.,][]{Mikami et al.(1992), Bachiller(1997)}.
Here, however, the SiO emission has a narrow line width about 1 \kms\ toward the {\bf arc}-like structure (Figure \ref{spectral} and \ref{channel_1}c).
The lack of a broad line width may be evidence that the shock is relatively old and that the turbulent motions produced within the shock region have dissipated.
Such a fossil shock has been suggested as the explanation for similar observations of the sources HH7-11 (SVS13-A) and NGC 2264 by \cite{Codella et al.(1999)} and \cite{Lopez-Sepulcre et al.(2016)}, respectively.
This hypothesis will be further discussed in relation to the origin of the arc-like structure in Section 5.2.
We note that narrow line width of SiO can also be interpreted in terms of magnetohydrodynamic C-shocks \citep{Jimenez-Serra et al.(2004), Jimenez-Serra et al.(2005), Jimenez-Serra et al.(2008), Jimenez-Serra et al.(2009)}.

\par \iras\ is located at the edge of Lupus-1 molecular cloud, as can be seen in large-scale dust continuum and CO observations \citep{Tothill et al.(2009), Gaczkowski et al.(2015), Mowat et al.(2017)}.
A protostellar source has never been found through infrared observations \citep{Rygl et al.(2013)}  toward the southeastern region of \iras\ where the arc-like structure resides.
Furthermore, sub-mm continuum emission is not detected at this location in our data sets, nor has ALMA previously uncovered continuum emission at this location \citep{Jorgensen et al.(2013), Oya et al.(2014), Okoda et al.(2018)}. 
Combined with the previously mentioned lack of significant \co\ emission, these null results suggest that there is no nearby enhancement of dense gas to the southeast of \iras.
Thus, the arc-like structure likely originates from the interaction of an outflow oriented along the northwest to southeast axis ejected from \iras\ and its natal ambient gas.
If this is the case, a detectable counter lobe might be expected to the northwest of \iras. A hint of this counter lobe can be seen in the SO emission, as described in the next subsection.

\subsection{Linear Structure along the Northwest to Southeast Direction}
\par
\begin{figure}[h!]
\centering
\includegraphics[scale=0.5]{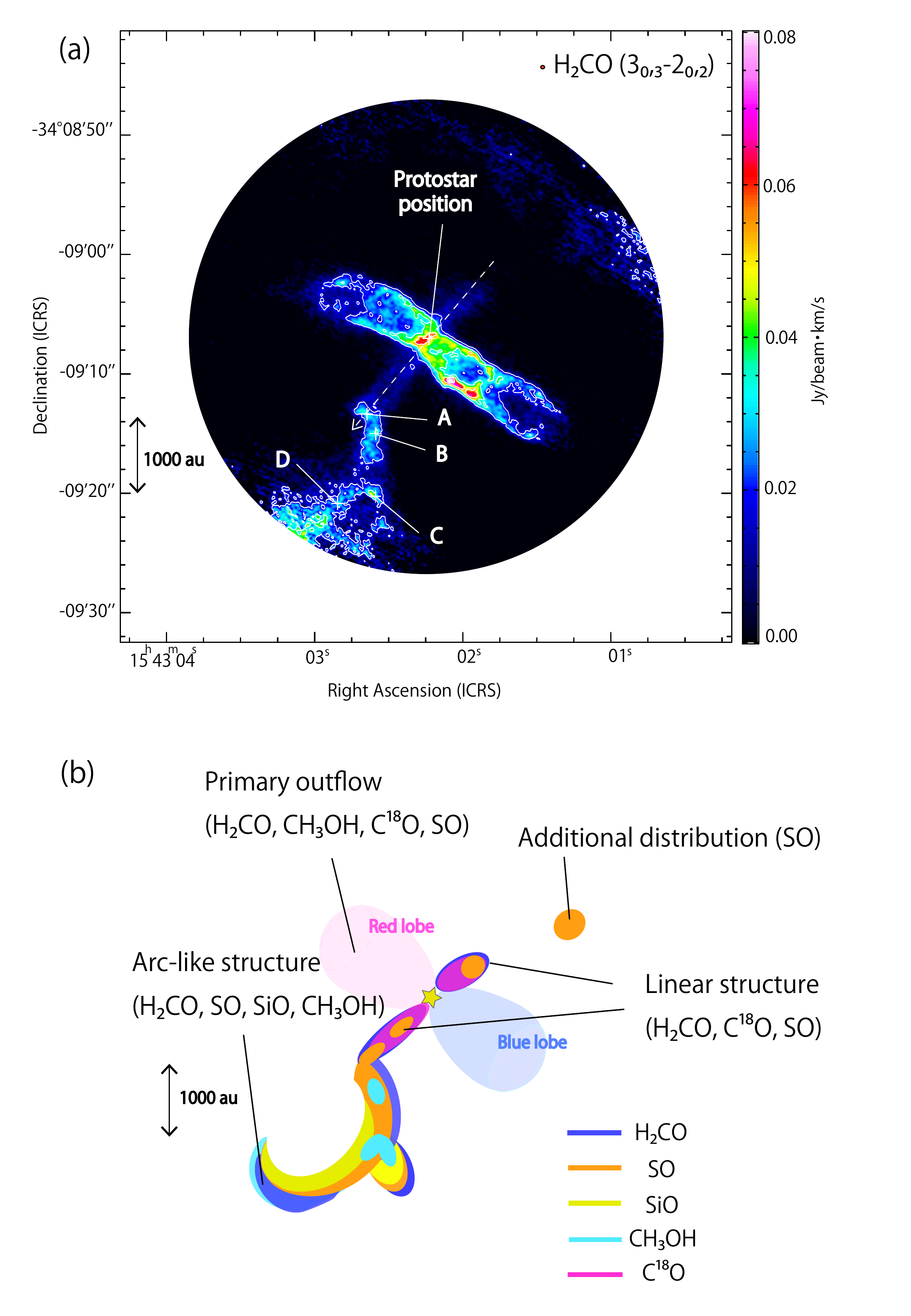}
\caption{(a) Moment 0 map of the \hhco\ line emission. Contour levels are every $\sigma$ from 3$\sigma$, where $\sigma$ is 6 \mjybeam. The line intensities are integrated from 3.7 \kms\ to 5.7 \kms\ to focus on the {\bf arc}-like feature and linear structure. White cross marks show the five locations where molecular line profiles are produced (See Figure \ref{spectral}). The red circle at top-right represents the synthesized beam size. 
The dashed arrow represents the direction of the PV diagram.
(b) Schematic picture of the molecular distributions. \label{moment_1}}
\end{figure}

\begin{figure}[h!]
\centering
\includegraphics[scale=0.4]{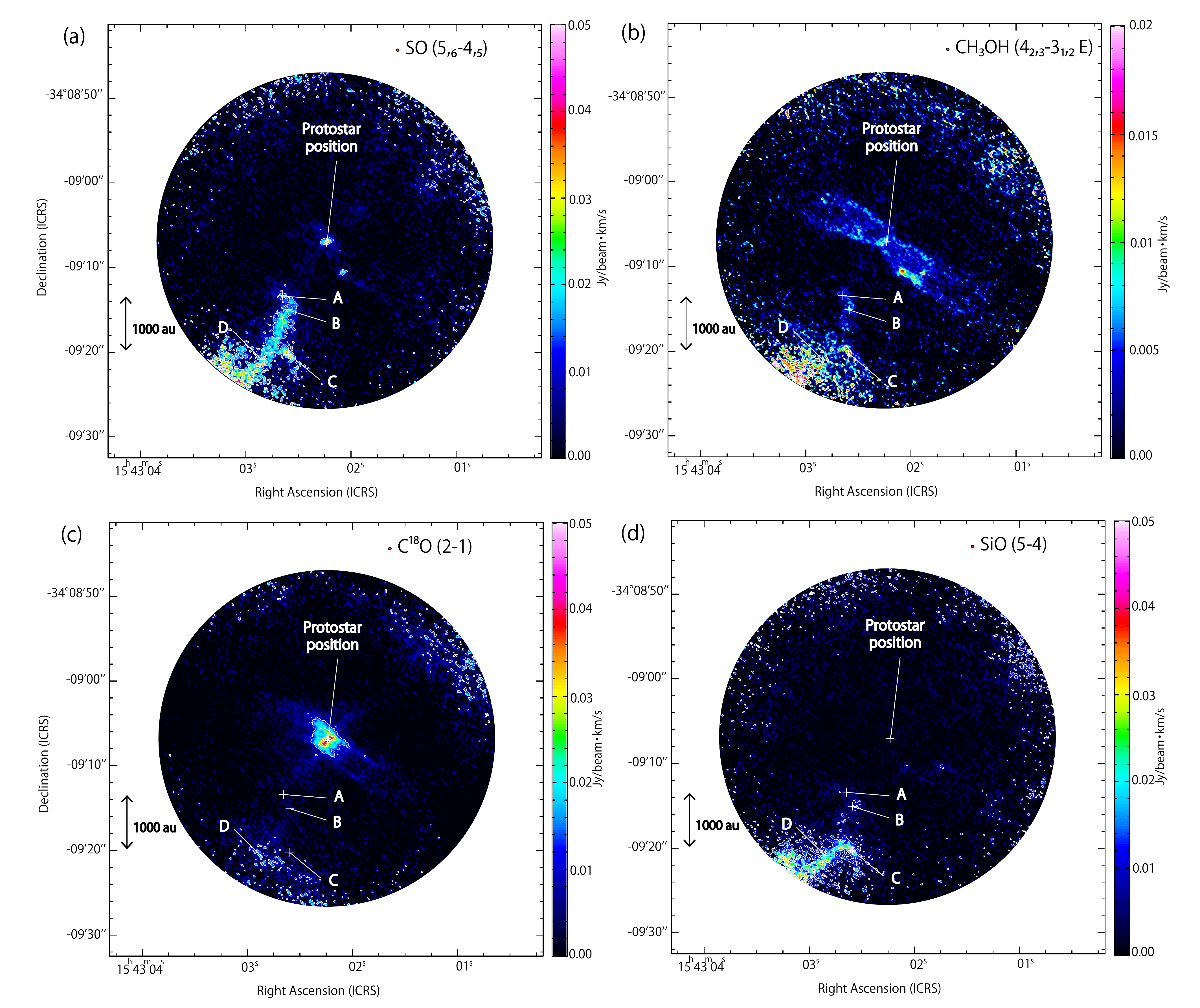}
\caption{(a,b,c,d) Moment 0 maps of the SO, \meta, \co\ and SiO line emission. Contour levels are every $\sigma$ from 3$\sigma$, where $\sigma$ is 4 \mjybeam\kms, 3 \mjybeam\kms, 4 \mjybeam\kms, and 3 \mjybeam\kms, respectively. See Figure \ref{moment_1} caption for additional details.\label{moment_2}}
\end{figure}

\begin{figure}[h!]
\centering
\includegraphics[scale=0.7]{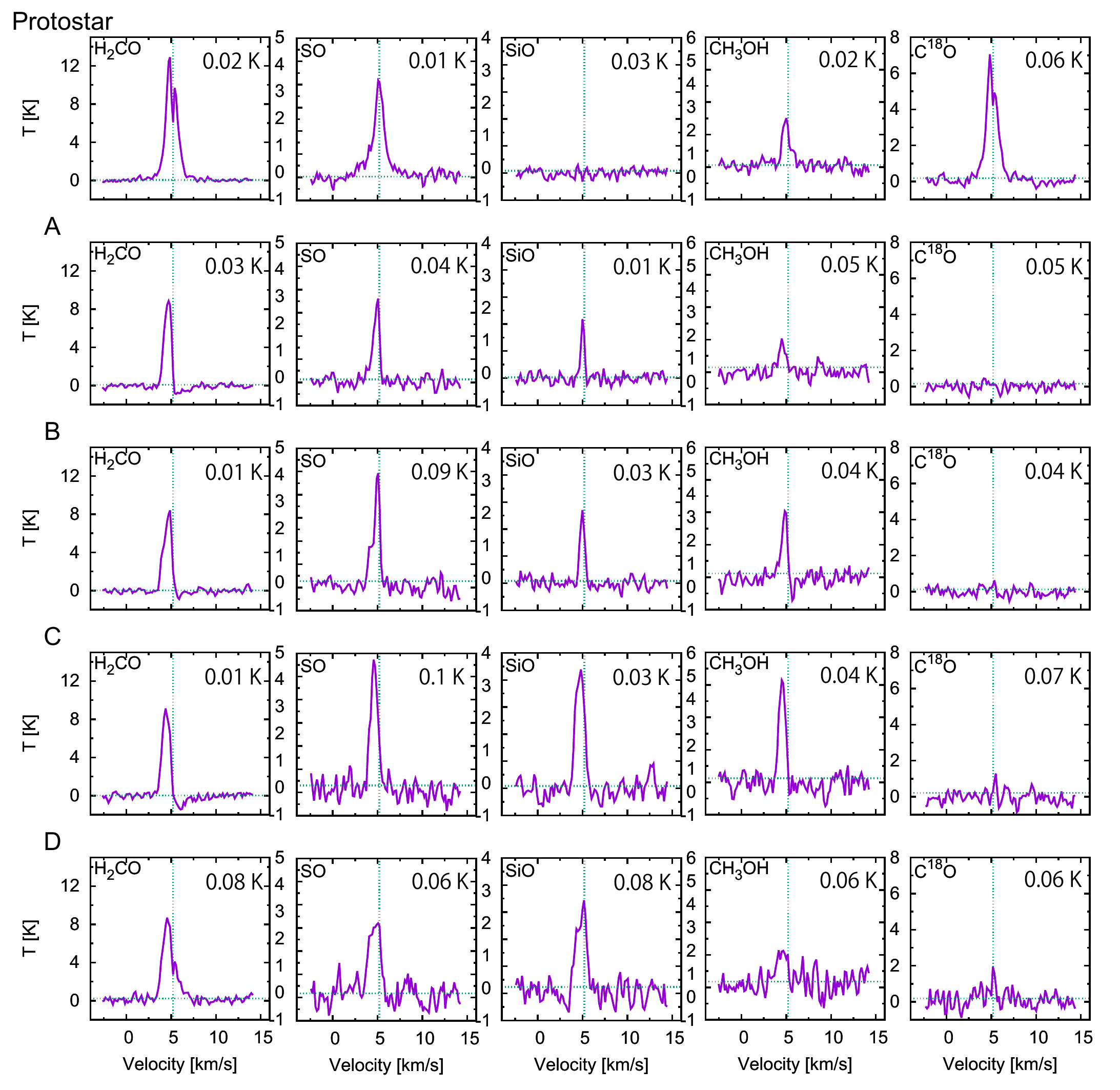}
\caption{Molecular line profiles observed toward the four positions (A-D; see Figures \ref{moment_1} and \ref{moment_2}) in the {\bf arc}-like structure and at the protostellar position. These spectra are extracted from the apertures of 1\farcs0. The horizontal green dashed lines represent each 3 $\sigma$.
The top-right values indicate $\sigma$ for each line profile.
The vertical green dashed lines represents the systemic velocity of the protostellar core, 5.2 \kms\ \citep{Yen et al.(2017)}\label{spectral}}
\end{figure}

 Focusing on the vicinity of the protostar, within 5$''$, a linear structure is seen in the \hhco\ emission, connecting the southeast arc-like structure to the protostar and continuing toward the northwest (Figure \ref{moment_1}a).
This structure is more clearly identified in the velocity maps at 4.7 \kms\ and 5.1 \kms\ (Figure \ref{channel_1}a).
Investigating this linear structure near the protostar, we find that the geometry is not simply an extension of the disk/envelope system related to the primary outflow (Figure \ref{shift}).
The linear structure is slightly inclined by about 10\degr\ from the disk/envelope axis (P.A. 130\degr) reported previously \citep{Oya et al.(2014), Yen et al.(2017), Okoda et al.(2018)}, and their origin ($\alpha$ and $\beta$ in Figure \ref{shift}a) are offset by 1$''-$2$''$ ($\sim$150$-$300 au) from the disk/envelope axis.
This feature is also seen weakly in \co\ emission over the same velocity range of 4.5 \kms\ to 5.7 \kms\ (Figure \ref{shift}b).

\par The SO emission also shows evidence of an extension to the northwest in the velocity maps of 5.1 \kms\ and 5.5 \kms\ (Figure \ref{channel_1}b).
This emission is extended from the protostar along the same orientation as the linear structure (P.A. 140\degr) in the 5.1 \kms\ channel map (Figure \ref{shift}).
At 5.5 \kms, additional emission appears further toward the northwest and shows a slight offset from the above axis.
In contrast to the emission from \hhco, SO is more structured and less continuous in the vicinity of the protostar. 
Nevertheless, the observed SO emission is consistent with the geometry seen in the \hhco\ emission.
The morphology near the protostar in the \hhco\ and \co\ emission and the apparent symmetry in the SO emission strongly suggest that these features are part of an outflow structure, rather than a structure formed from gas accreting toward the protostar as seen in the source Per-emb-2 by \cite{Pineda et al.(2020)}.

\par The possible secondary outflow is verified by the velocity structure.
Figure \ref{pv}a presents the position-velocity (PV) diagram for the \hhco\ emission along the northwest to southeast axis (P.A. 140\degr), centered at the protostellar position.
The position axis is approximately along the linear feature seen in the \hhco\ emission (Figure \ref{moment_1}a).
The velocity toward the northwest remains roughy constant and only slightly blue-shifted with respect to the systemic velocity.
In contrast, toward the southeast there is a clear velocity gradient, with the emission becoming more blue-shifted with increasing distance from the protostar and toward the arc-like structure.
Furthermore, as seen in the PV diagram, the velocity width increases up to about 1 \kms\ at the southeastern terminus (Position A; Figure \ref{moment_1}a). 
This location is coincident with the northern tip of the arc-like structure.
A similar velocity structure can be seen in the PV diagram for the \co\ emission (Figure \ref{pv}b).
The velocity gradient is calculated to be about 1.2 \kms\ over 1200 au, assuming that the systemic velocity is 5.2 \kms.
This gradient is unexpectedly large if identified as infalling motion to the protostar.
A more natural explanation is that the velocity gradient is associated with outflow motion.
A lack of a red-shifted component in the northwestern part is puzzling but may be due to less gas being affected by the outflow.
Alternatively, this could be reproduced if accretion of material is not symmetric \citep{Zhao et al.(2018)}.
The slight shift of the origin points of the linear structure near the protostar (Figure \ref{shift}) can also be understood if these origin points are part of this secondary outflow.
\par One may expect that the secondary outflow would be detected in the $^{12}$CO lines, which are the most commonly observed outflow tracers. 
Indeed, a faint blue-shifted component near the arc-like structure on the southeastern part can marginally be seen in the interferometric $^{12}$CO ($J=2-$1) observations by \cite{Bjerkeli et al.(2016a)} and \cite {Yen et al.(2017)}.
This would be a remnant of part of the arc-like structure.
On the other hand, the linear structure and its extension to the northwest are not seen in the $^{12}$CO emission.
The velocity shift is so close to that of the ambient gas that they are likely to be resolved out. The absence of the red-shifted emission at the northwestern part is puzzling, as mentioned above.

\section{Analyses of the Secondary Outflow}\label{ana}
\subsection{Dynamical Timescale of the Outflow and Outflow Mass}
We have identified a linear structure along the northwest to southeast axis (P.A. 140\degr) as well as an arc-like structure further to the southeast.
A northwestern counterpart can marginally be seen in SO emission.
This extended feature is most likely a secondary outflow, which has not been previously identified. 
Collimated features similar to the linear structure seen near the protostar are expected to form around young protostellar sources \citep*[e.g.,][]{Machida et al.(2008), Inutsuka(2012), Velusamy et al.(2014), Busch et al.(2020)}.


\par We evaluate the dynamical timescale of the secondary outflow from the PV diagram of the \hhco\ emission (Figure \ref{pv}a).
The plane of the sky distance and the line-of-sight velocity shift from the systemic velocity are 1200 au and $\sim$1.2 \kms, respectively. Thus, the timescale is estimated to be $\sim5000\times \cot(i)$ yr, where $i$ is the inclination angle of the outflow axis ($i=$0\degr\ for pole-on).
Considering the morphology of the observed secondary outflow, the axis may be close to the plane of the sky ($i=$90\degr) and the dynamical timescale is regarded as an upper limit.
The observed feature might also be a relic of a previous fast outflow, considering the relatively narrow observed  line width (See Section 5.2).
In this case, the observed velocities might be significantly lower than they were in the past, and the calculated dynamical timescale would be an overestimate. 
Note that the dynamical timescale of the primary outflow (P.A. 220 \degr) is reported to be 10$^{2}-10^{3}$ yr \citep{Oya et al.(2014), Yildiz et al.(2015), Bjerkeli et al.(2016a)}.
Hence, the secondary outflow would likely have formed before the primary outflow.
The dynamical timescale of the secondary outflow is comparable to the time for depletion of SiO onto dust grains: 10$^3-10^4$ yr if the H$_2$ density is roughly 10$^{5}-10^{6}$ cm$^{-3}$ \citep*[e.g.,][]{Caselli et al.(1999)}.
Thus, if SiO molecules were liberated from dust grains in the past by shocks related to a secondary outflow, they would be able to survive in the gas phase until the present day.
We roughly estimate the total mass within the observed secondary structure seen in SO emission to be 10$^{-5}$$-$10$^{-4}$ \sm, where we use the apparent size and an assumed H$_2$ density of 10$^{5}-10^{6}$ cm$^{-3}$ based on the critical densities of the observed lines.
Similarly, we estimate the mass of the primary outflow from the \hhco\ emission to be 10$^{-4}$$-$10$^{-3}$ \sm, assuming the same H$_2$ density range.

\subsection{Molecular Abundances around the Arc-like Structure}
\par 
In order to compare the molecular abundances of the \iras\ with other outflow shock regions, we derive the lower limits to the abundance ratio for several molecules relative to \co\ at the four positions (A-D) and the abundance ratio at the protostar position (Figures \ref{moment_1} and \ref{moment_2}).
Positions A and D are the bending point from the linear structure toward the arc-like structure and the south of the arc-like structure, respectively, while positions B and C are the local peak positions of \meta\ in Figure \ref{moment_2}b.
We calculate the column densities of \hhco, SO, SiO, and \meta\ as well as the upper limit of \co\ by using the non-LTE radiative transfer code, RADEX\footnote{The collisional rates of \hhco, SO, SiO, \meta, and \co\ are originally calculated by \cite{Wiesenfeld and Faure(2013)}, \cite{Lique et al.(2006)}, \cite{Balanca et al.(2018)}, \cite{Rabli and Flower(2010)}, and \cite{Yang et al.(2010)}, respectevely.} \citep{van der Tak et al.(2007)}, toward these positions, assuming that the gas temperature is 20$-$80 K and the H$_2$ density is 10$^5-10^6$ cm$^{-3}$ (Table \ref{column}).
Regarding the \co\ column density at the four position, we use the upper limits to the intensity and the mean line width of other molecular lines (\hhco, SO, SiO, and \meta).
Any extended component traced by the \co\ emission at the systemic velocity may be resolved out. 
In this case, the upper limits to the \co\ intensity might be {\bf underestimated}.
However, we here discuss the molecular abundances in the arc-like structure which is blue-shifted from the systemic velocity by 0.5 \kms.
Such a compact structure should be observed even in the \co\ emission without the resolved-out problem.
Moreover, the maximum recoverable size (12\farcs8) is larger than the arc-like structure (at most 10$''$).
For these reasons, the resolved-out effect should not affect the molecular abundances significantly.
For the protostar position, the \co\ column density is derived to be (5.5$-$6.4)$\times10^{15}$ cm$^{-2}$, where the gas temperature and the H$_2$ density are assumed to be 54 K \citep{Okoda et al.(2020)} and 10$^5-10^7$ cm$^{-3}$, respectively.
Then, the abundance ratios relative to \co\ are evaluated from the column densities, where the same temperature is assumed for the molecular species and \co.
\par Only lower limits are obtained for the abundance ratios at all the four positions.
Nevertheless, the abundances of SO, SiO, and \meta\ are found to be significantly enhanced in the shocked region in comparison with the prostostar position (Table \ref{ratio}). 
Moreover, the abundance ratios are as high as those reported for L1157-B1 (Table \ref{ratio}), further supporting the outflow shock.
We note, however, that the physical conditions of L1157-B1 are likely different from those of \iras. L1157-B1 shows a broad line width of SiO.
Hence, we also compare the SiO abundance of the {\bf arc}-like structure to that of the shocked region with the narrow SiO line in the Class I protostellar source, HH7-11 (SVS13-A) (\cite{Codella et al.(1999)}; $t_{des}\sim$10$^4$ yr).
We estimate the SiO abundance relative to H$_2$ by using the nominal \co\ fractional abundance of 1.7$\times10^{-7}$ \citep{Frerking et al.(1982)}.
In Table \ref{ratio}, the abundance of the arc-like structure is much higher than that found for HH7-11 (SVS13-A).
This result seems to be related to the shorter timescale of the secondary outflow than that of the HH7-11 (SVS13-A) outflow, $\sim10^5$ yr \citep{Lefloch et al.(1998)}.

\section{Origin of the Secondary Outflow}\label{sec-disc}

\subsection{Scenario 1: Two Outflows Driven by a Binary System}
It is well known that some Class 0 protostellar sources contain a binary or multiple system \citep*[e.g.,][]{Tobin et al.(2016),Tobin et al.(2018)}.
A possible explanation for the secondary outflow is that \iras\ is a binary system, launching outflows in two different directions with respect to the plane of the sky. 
Such binary/multiple systems launching more than one distinct outflow toward different directions have been previously observed within IRAS 16293-2422 Source A \citep{van der Wiel et al.(2019), Maureira et al.(2020)}, BHR71 \citep{Zapata et al.(2018), Tobin et al.(2019)}, NGC1333 IRAS2A \citep{Tobin et al.(2015)}, and NGC2264 CMM3 \citep{Watanabe et al.(2017)}.
In all these cases, however, the binaries are separated by more than 40$-$50 au.
Most binary sources observed to have a circumbinary disk/envelope structure show only a single outflow or parallel outflows, e.g., BHB07-11 \citep{Alves et al.(2017)}, L1448 IRS3B-a, b \citep{Tobin et al.(2016)}, and L1551NE \citep{Reipurth et al.(2000), Lim et al.(2016)}. 
An exception is the VLA1623 case.
\cite{Hara et al.(2020)} report that this source is a close binary system with a separation of 34 au, and two molecular outflows are inclined by 70\degr\ from each other across the plane of the sky.
\par High-resolution observations ($\sim$30 au) of \iras\ with ALMA  by \cite{Okoda et al.(2018)} show a single peak in the dust continuum, indicating that this source is not a wide binary.
In addition, they find a well defined Keplerian disk structure with a size of 40 au in the SO line. 
If \iras\ is a close binary system instead of being two independent systems that should be accidentally aligned in a line of sight of our observation, it poses a very difficult question on how to create such complex system from a molecular cloud core. 
In principle, a single molecular cloud core can have very complicated internal angular momentum distribution, but there is no theory to explain the formation of binary stars with an apparent separation much smaller than the above ALMA resolution ($\sim$30 au) and with very different rotation axes.
Indeed, MHD simulations by \cite{Matsumoto et al.(2017)} reveal that a single outflow is launched in the case of complex angular momentum distribution in the core.
Note also that it is very unlikely that two, almost orthogonal, outflows would be launched from the circumbinary disk.
We thus conclude that binary hypothesis for the secondary outflow is unlikely, although we cannot completely rule out the possibility of a tight binary ($<$30 au) system or that of a binary system whose components are aligned close to the line of sight. 
More observations at higher resolution will be required to investigate the multiplicity of \iras.

\subsection{Scenario 2: Outflow Outburst and Reorientation}
\par A second possible explanation for the observed secondary structure is a past reorientation of the outflow launched from a single protostar. 
Assuming that outflow reorientation likelihood is random, the observed orthogonality of the secondary and primary outflows suggest a low expectation for this specific geometry, especially if the secondary outflow lies nearly along the plane of the sky as in the case of the primary outflow.
This difficulty can be mitigated if the secondary outflow is a relic and the fast-moving component expected along the outflow has already been dissipated.
In this case, the observed outflow direction may be significantly inclined with respect to the plane of the sky and the observed velocity may be close to the true velocity of the present relic.
\par The dissipation timescale of the high velocity component can be roughly calculated as \citep{Codella et al.(1999)}:
\begin{equation}
t_{dis}=\frac{8}{3}\frac{\rho}{\rho{_1}}\frac{r}{v},
\end{equation}
where $\rho/\rho{_1}$ is the mass density ratio of the shocked gas over the ambient gas, $r$ the size of the shocked region, and $v$ the shock velocity.
In the case of \iras, we assume the shock velocity to be 10 \kms\ and employ the size of the shocked region of 300 au according to a typical width of the arc-like structure (Figure \ref{moment_1}).
In addition, the $\rho/\rho{_1}$ ratio is assumed to be 10.
The turbulent dissipation timescale is estimated to be $\sim$4000 yr based on the above assumptions.
This timescale is comparable to the dynamical timescale of the secondary outflow ($\sim5000\times \cot(i)$ yr).
Thus, the absence of the high-velocity component seems reasonable. 

\par It is further worth pointing out two specific features of \iras\ in relation to the reorientation hypothesis.
First, the protostellar mass is reported to be 0.007 \sm\ on the basis of the disk Keplerian motion \citep{Okoda et al.(2018)}.
This mass is lower than the expected mass of the first hydrostatic core proposed by star formation theories \citep*[0.01$-$0.05 \sm;][]{Larson(1969), Masunaga et al.(1998), Saigo&Tomisaka(2006), Inutsuka(2012)}.
Thus, \iras\ appears to be in the earliest stages of protostellar evolution, a notion reinforced by the significantly larger reported envelope mass, 0.5$-$1.2 \sm \citep{Kristensen et al.(2012), Jorgensen et al.(2013)}.
Second, past episodic accretion events have been suggested for this source by \cite{Jorgensen et al.(2013)} and \cite{Bjerkeli et al.(2016b)}.
\cite{Jorgensen et al.(2013)} find a ring structure of H$^{13}$CO$^+$ at a scale of 150$-$200 au around the protostar. The lack of H$^{13}$CO$^+$ interior to the ring is inconsistent with the present heating rate from the central protostar; however, a previous burst of enhanced luminosity due to an accretion burst would have removed that H$^{13}$CO$^+$ through chemical reaction with sublimated H$_2$O from dust grains.
The authors predict that the accretion burst occurred $10^2-10^3$ yr ago, consistent with the dynamical timescale of the primary outflow \citep{Oya et al.(2014), Bjerkeli et al.(2016a)}. 
\cite{Bjerkeli et al.(2016b)} report that the HDO ($1_{0,1}-0_{0,0}$) emission is localized on the cavity wall in the vicinity of the protostar, interpreting that this is also due to a past accretion burst. 
The timescale for the outflow reorientation would be similar to the interval of episodic accretion in the case that the angular momentum of the gas in a molecular core is non-uniform (See below).

\par Assuming that the secondary structure reflects a previous change in the outflow direction for \iras, can the observed velocity and orientation of the feature be reconciled?
Relatively high velocity shocks ($>$ 5$-$20 \kms) are required to liberate SiO from dust grains \citep*[e.g.,][]{Caselli et al.(1997), Jimenez-Serra et al.(2008)}.
Thus, the low, $\sim 1$ \kms\ velocity shift of the SiO arc-like structure needs to be considered carefully.
\par As discussed above, the possibility that the secondary outflow velocity is underestimated due to the structure lying very close to the plane of the sky is small. 
Alternatively, a more likely hypothesis is that the observed secondary outflow represents a relic structure which is no longer powered by current mass ejection from the protostar and inner disk.
Thus, the fast component of the secondary outflow should have dissipated through interaction with ambient gas leaving only the observed motions close to the systemic velocity.
This is also the reason why we can see the linear structure from the vicinity of the protostar.
Previous observations of low-velocity features in SiO have also suggested this hypothesis \citep{Lefloch et al.(1998), Codella et al.(1999), Lopez-Sepulcre et al.(2016)}.
The relatively narrow line widths observed for the shock tracer lines described in Section 3.2 are consistent with this picture.

\par We must also reconcile the orientation of the secondary outflow, which is almost perpendicular to the angular momentum axis of the currently accreting gas. 
If the angular momentum of the episodically accreting gas varies with time, the direction of the outflow axis may change drastically, even as much as observed in this source. 
According to \cite{Misugi et al.(2019)}, the angular momentum of the gas in a molecular core is related to the degree of centroid velocity fluctuations within the parental filamentary molecular cloud, resulting in the angular momentum versus cloud mass relation determined by the Kolmogorov power spectrum of weak (i.e. subsonic or transonic) turbulence. 
In this analysis, the total angular momentum of a star-forming core is a vector sum of the various angular momenta of turbulent fluid elements whose directions are almost randomly oriented. 
The summation over angular momentum vectors inevitably includes cancellation of opposite components of angular momentum vectors.

\par The identification of such an origin for the core rotation has important implications, because the actual collapse process of the core is not homologous but rather undergoes a $''$run-away$''$ process where a central dense region collapses first and the outer regions accrete onto the central region later. Thus, episodic accretion events with very different angular momenta are expected to occur during this run-away collapse process.  
Applying this model to the early evolution of the \iras\ protostellar core, and recognizing that the mass of the central protostar is extremely low, there should only have been a very few discrete episodic accretion events and random angular momentum reorientations. 
Moreover, \cite{Okoda et al.(2018)} report that the central velocity of the disk structure observed with the SO line is 5.5 \kms, which is shifted 0.3 \kms\ from the systemic velocity of the protostellar core at 5.2 \kms\ \citep{Yen et al.(2017)}. 
This previous result may also naturally caused by the continual accretion of fluid elements that have various momenta as well as various angular momenta.
The observed secondary outflow feature provides plausible observational evidence for this picture. 
It is interesting to note that a change in the angular momentum axis of the accreting gas has also recently been suggested observationally for some other young protostellar sources \citep{Sakai et al.(2019), Zhang et al.(2019), Gaudel et al.(2020)}.
\par If the rotation axis is misaligned with the global magnetic field direction, the MHD simulations show that the directions of the jet/outflow and disk randomly change over the time \citep{Matsumoto(2004), Hirano et al.(2020), Machida et al.(2020)}.
Since the misalignment can naturally be caused by the above picture, this mechanism would also contribute to the outflow reorientation of this source.
Note that the global direction of the magnetic field around \iras\ is almost along the primary outflow direction \citep{Redaelli et al.(2019)}.

\subsection{Future Directions}
\par Unfortunately, the above interesting theoretical process is not yet well theoretically studied because most of the numerical MHD simulations for the formation of protostars and disks have been performed with the simplest initial condition for the rotation, i.e., rigid body rotation throughout the core \citep*[e.g.,][]{Machida & Basu(2019), Inutsuka(2012), Tsukamoto(2016), Wurster&Li(2018)}.
Such turbulent motion, however, could be important for angular momentum variation within the core as well as the thermal pressure of the gas. 
\par We propose the reorientation scenario as one possibility to account for this observation.
To understand the interesting feature of the secondary outflow, we will need additional observations to investigate other possibilities carefully, particularly, the existence of the close binary system ($<$30 au).
For testing the close binary hypothesis, centimeter-wave observations at a high angular resolution ($<$30 au) are essential to resolve the components without obscuration from the optically thick dust continuum. 
Mosaic observations with ACA would be useful to explore the environment around this source  in detail. 
It is also interesting to observe the velocity field of the parent core at a high angular resolution in order to investigate the distribution of the angular momentum as previously done at a larger scale for the other sources \citep*[e.g.,][]{Caselli et al.(2002)}.
Furthermore, observations in other molecular lines (CCH, CS, \ccchh, N$_2$H$^+$, and \hthcop\ etc) would promote our understanding of the newly found structures, which are now in progress in the FAUST program.


 \section{Conclusions}
 \label{sec-conc}
 \par We have uncovered an interesting feature along the direction almost perpendicular to the primary outflow in \iras.
 This is most likely the relic of a secondary outflow ejected from the single protostar at the center of this system. 
 The observational results of this paper present an important implication on the earliest stage of star formation.
 The main results are listed below.
\par 1.  We have identified an arc-like structure in SO, SiO, and \meta\ line emission.
 Since these molecular species are known to be shock tracers, the arc-like structure is most likely a shock region produced by a relic outflow.
 The molecular abundances of \hhco, SO, SiO, and \meta\ in the arc-like structure are clearly enhanced in comparison with those at the protostar position, and they are consistent with those reported for shocked regions in the source L1157-B1.

\par 2. The \hhco, SO and \co\ line emission produce a linear feature around the protostar. 
The SO emission also reveals  an additional structure to the northwest.
In the PV diagram of the \hhco\ and \co\ emission, the velocity increases with increasing distance from the protostar toward the southeast.
Such morphological and kinematic features are consistent with those expected from an outflow.

\par 3. We roughly estimate the dynamical timescale of the secondary outflow to be $\sim5000\times \cot(i)$ yr ($i=$0\degr\ for pole-on).
The timescale is similar to that of depletion of SiO onto dust grains, which is 10$^3-10^4$ yr.

\par 4.  As previously reported by \cite{Okoda et al.(2018)}, the Keplerian disk structure around \iras\ is seen in the SO line emission at a resolution of 30 au, and the dust continuum maps show only a single peak.
Given this geometry, the launch of two almost orthogonal outflows from the circumbinary disk ($<$30 au) is very unlikely, although we cannot completely exclude a tight binary hypothesis. Higher resolution observations are needed to examine the possibility further.

\par 5.  We thus hypothesize that the secondary outflow is a relic of a past reorientation of the outflow launched from a single protostar.
The narrow line width of the SiO emission implies that the arc-like structure has dissipated the turbulent motions associated with the earlier shocks.
The change in the direction of the outflow axis may be related to non-uniform internal angular momentum distribution in the molecular core, advected onto the central region via episodic accretion. Such events may occur during the earliest stages of protostar formation like \iras.

\acknowledgments
\par The authors thank the anonymous reviewer for invaluable comments.
They are also grateful to Hauyu Baobab Liu for useful discussions.
This paper makes use of the following ALMA data set:
ADS/JAO.ALMA\# 2018.1.01205.L (PI: Satoshi Yamamoto). ALMA is a partnership of the ESO (representing its member states), the NSF (USA) and NINS (Japan), together with the NRC (Canada) and the NSC and ASIAA (Taiwan), in cooperation with the Republic of Chile.
The Joint ALMA Observatory is operated by the ESO, the AUI/NRAO, and the NAOJ. The authors thank to the ALMA staff for their excellent support.
This work is supported by the European Research Council  (ERC) under the European Union's Horizon 2020 research and innovation programmes: $''$The Dawn of Organic Chemistry$''$ (DOC), grant agreement No 741002, and “Astro-Chemistry Origins” (ACO), Grant No 811312.
The National Radio Astronomy Observatory is a facility of the National Science Foundation operated under cooperative agreement by Associated Universities, Inc.
I.J.-S. has received partial support from the Spanish FEDER (project number ESP2017-86582-C4-1-R) and the State Research Agency (AEI; project number PID2019-105552RB-C41).
D.J. is supported by NRC Canada and by an NSERC Discovery Grant. 
This project is also supported by a Grant-in-Aid from Japan Society for the Promotion of Science (KAKENHI: Nos. 18H05222, 19H05069, 19K14753).
Yuki Okoda thanks the Advanced Leading Graduate Course for Photon Science (ALPS) and Japan Society for the Promotion of Science (JSPS) for financial support.

{}

\begin{longrotatetable}
\begin{table}[ht]
\centering
\caption{Observation Parameters \label{observations}}
\scalebox{1.0}{
\begin{tabular}{llll}
\hline \hline
Parameter 					& Band 6 (C43-5) 							& Band 6 (C43-2) 									& 	Band 6 (7 m Array of ACA)\\
\hline
Observation date(s) 			&  2018 Oct 23							& 2019 Jan 06 									&2018 Oct 24, Oct 25 \\
Time on Source (minute)	& 47.10 								&12.63												&28.75, 28.77\\
Number of antennas			& 45 								&47											&10\\
Primary beam width  (\farcs)	& 26.7 								&26.7												&45.8\\										
Total bandwidth (GHz) 		&0.059								&0.059											&0.062\\
Continuum bandwidth (GHz)	&1.875								&1.875											&2.000\\
Proj. baseline range (m) 		&15.03$-$1310.74 				&12.66$-$425.02							      &7.43$-$46.74\\
Bandpass calibrator			&J1427-4206						&J1427-4206									&J1229+0203, J1427-4206\\
Phase calibrator				&J1626-2951 						&J1517-2422									& J1517-2422	\\
Flux calibrator				&J1427-4206						&J1427-4206				&J1229+0203, J1427-4206	\\
Pointing calibrator			&J1650-2943, J1427-4206			&J1427-4206									&J1229+0203, J1427-4206, J1517-2422\\
Resolution(\farcs$\times$\farcs(P.A.$^{\circ}$))&0.338$\times$0.280 (63.9)&1.140$\times$0.921 (79.9)&8.210$\times$4.780 (69.2)\\
Rms (mJy beam$^{-1}$channel$^{-1}$)			&1.8							&4.4									& 24.0\\
\hline
\end{tabular}}
\begin{flushleft}
\end{flushleft}
\end{table}
\end{longrotatetable}

\begin{table}[ht]
\caption{ List of Observed Lines $^a$ \label{line}}
\scalebox{0.9}{
\begin{tabular}{ccccccc}
\hline \hline
 Molecule&Transition & Frequency\ (GHz) & $S \mu^2$($D^2$) & $E_{\rm u}$$k^{-1}(\rm K)$& Beam size\\
 \hline
\hhco$^b$ &$ 3_{0,3}-2_{0,2}$	&      218.2221920    &16.308    & 21  &0\farcs35$\times$0\farcs30 (P.A. 69$^{\circ}$)\\   
SO$^c$ &$5_6-4_5$&   219.9494420    &14.015   &  35 &               0\farcs35$\times$0\farcs30 (P.A. 65$^{\circ}$) \\
SiO$^d$ &	5-4 &        217.1049190 &   47.993   &  31   &0\farcs35$\times$0\farcs30 (P.A. 68$^{\circ}$)\\	            
\meta$^e$& $4_{2,3}-3_{1,2}$   &     218.4400630  & 13.905   &  45      &0\farcs35$\times$0\farcs29 (P.A. 69$^{\circ}$) \\
\co$^f$ &$ 2-1$    &       219.5603541   & 0.024399 &  16           &0\farcs35$\times$0\farcs29  (P.A. 65$^{\circ}$)\\
\hline
\end{tabular}}
\begin{flushleft}
\tablecomments{
$^a$ Line parameters are taken from CDMS \citep{Endres et al.(2016)}.
 The beam size for each line is obtained from the observations.
 $^{b}$ \cite{Muller(2017)}
 $^{c}$ \cite{Klaus et al.(1996)}
 $^{d}$ \cite{Muller et al.(2013)}
 $^{e}$ \cite{Xu et al.(2008)}
 $^{f}$ \cite{Klapper et al.(2001)}
}
\end{flushleft}
\end{table}

\begin{table}[ht]
\centering
\caption{Column Densities [10$^{14}$ cm$^{-2}$] \label{column}}
\scalebox{0.9}{
\begin{tabular}{cccccc}
\hline \hline
Molecule &Protostar$^a$&A $^b$& B$^b$& C $^b$& D $^b$\\
\hline
\hhco\    & 0.4$-$2 & 0.2$-$2 & 0.2$-$2 & 0.2$-$3 & 0.2$-$3\\
SO		& 0.4$-$1 & 0.2$-$ 2 & 0.3$-$3 & 0.5$-$5 & 0.4$-$3\\
SiO	&$<$0.03$^c$ & 0.01$-$0.8 &0.02$-$1 &0.07$-$5 &0.05$-$3\\
\meta\	& 0.7$-$6 & 1$-$6 & 2$-$9 & 3$-$20& 2$-$10\\
\co & 55$-$64 & $<$1$^c$&$<$0.8$^c$ & $<$2$^c$ & $<$2$^c$\\
\hline
\end{tabular}
}
\begin{flushleft}
\tablecomments{
$^a$ The positions are shown in Figures \ref{moment_1} and \ref{moment_2}.
The temperature and the H$_2$ density are assumed to be 54 K \citep{Okoda et al.(2020)} and 10$^5-10^7$ cm$^{-3}$, respectively.
A range of the column density is shown for each molecule.\\
$^b$ The temperature and the H$_2$ density are assumed to be 20$-$80 K and 10$^5-10^6$ cm$^{-3}$, respectively.
A range of the column density is shown for each molecule at each position.\\
$^c$ The 3 $\sigma$ upper limits, where $\sigma$ is shown in Figure \ref{spectral}.
}
\end{flushleft}
\end{table}

\begin{table}[ht]
\centering
\caption{Molecular relative abundance ratios with respect to the \co\ emission \label{ratio}}
\scalebox{0.9}{
\begin{tabular}{cccccccc}
\hline \hline
Molecule &Protostar$^a$&A$^a$ & B$^a$& C$^a$ & D$^a$ & L1157-B1$^b$ & HH7-11 (SVS13-A)\\
\hline
$\lbrack$\hhco$\rbrack$/$\lbrack$\co$\rbrack$	&0.005$-$0.03&$>$ 0.2&	$>$ 0.2&	$>$ 0.1&	$>$ 0.1	& 0.5$-$1&$-$ \\
$\lbrack$SO$\rbrack$/$\lbrack$\co$\rbrack$		&0.007$-$0.02&$>$ 0.2&	$>$ 0.4&$>$	0.3	&$>$ 0.2	& 0.5$-$0.8&$-$\\
$\lbrack$SiO$\rbrack$/$\lbrack$\co$\rbrack$		&$<$0.001	&$>$ 0.01 &	$>$ 0.03	&$>$ 0.04	&$>$ 0.02	&0.1 &$-$\\
$\lbrack$\meta$\rbrack$/$\lbrack$\co$\rbrack$	&0.01$-$0.1&$>$ 1 &	$>$	2	&$>$	2 &$>$	1	&0.8$-$4&$-$\\
 X(SiO)$^c$[10$^{-9}$]& & 1.7 & 5.1 & 6.8& 3.4 & 17 & 0.08$-$0.3$^d$\\
\hline
\end{tabular}
}
\begin{flushleft}
\tablecomments{
$^a$ These values are derived by assuming the same temperature for the two molecular species for the ratio.\\
$^b$ The abundance ratios in the outflow-shocked region of L1157-B1. 
Values are extracted from \cite{Bachiller(1997)}.\\
$^c$ X(SiO) indicates the SiO abundance relative to H$_2$ estimated by using the normal \co\ abundance, 1.7$\times10^{-7}$ \citep{Frerking et al.(1982)}.\\
$^d$ These values are repoted in \cite{Lefloch et al.(1998), Codella et al.(1999)}.
}
\end{flushleft}
\end{table}

\begin{figure}[h!]
\rotatebox{90}{
\begin{minipage}{\textheight}\centering
\includegraphics[scale=0.85]{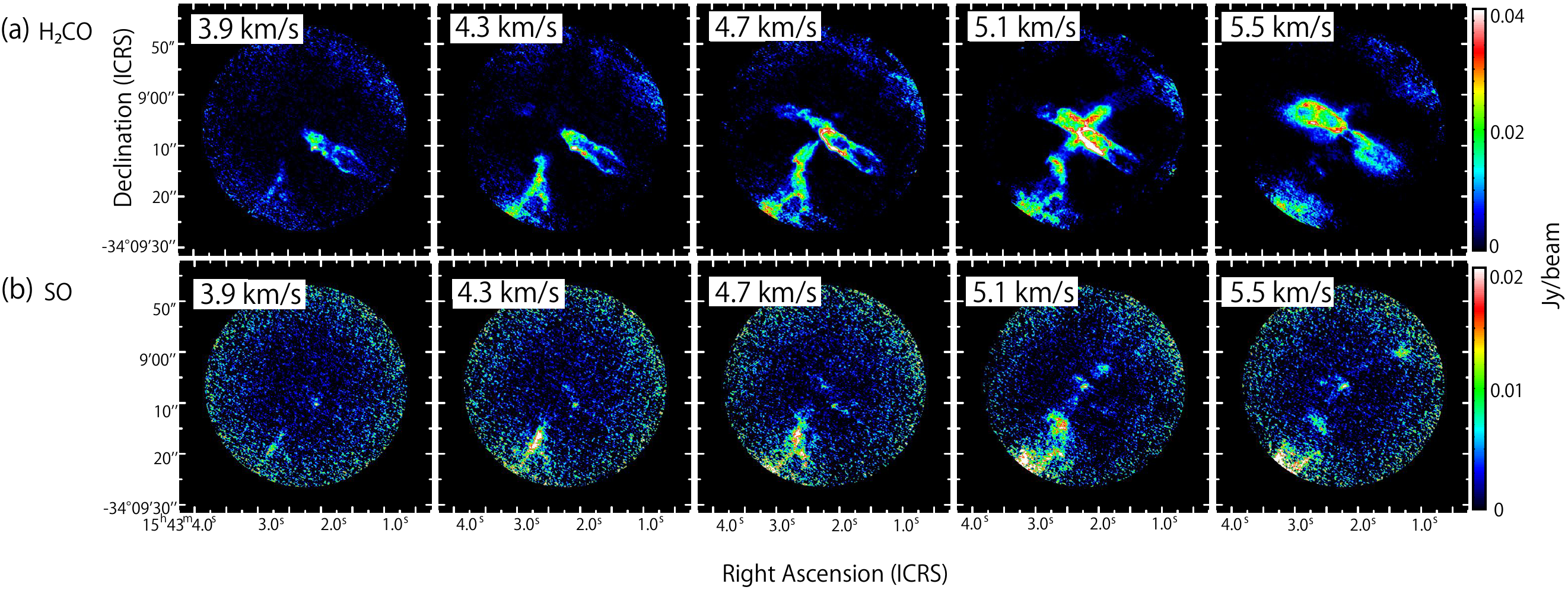}
\caption{Velocity channel maps of SO, \hhco, SiO, \meta, and \co.
Each panel represents the intensity averaged over a velocity range of 0.4 \kms\ centered at the quoted velocity.
For instance, the 3.9 \kms\ panel shows the average between 3.7 \kms\ and 4.1 \kms.
The systemic velocity of the protostellar core is 5.2 \kms.\label{channel_1}}
\end{minipage}}
\end{figure}
\addtocounter{figure}{-1}
\begin{figure}[h!]
\rotatebox{90}{
\begin{minipage}{\textheight}\centering
\includegraphics[scale=0.85]{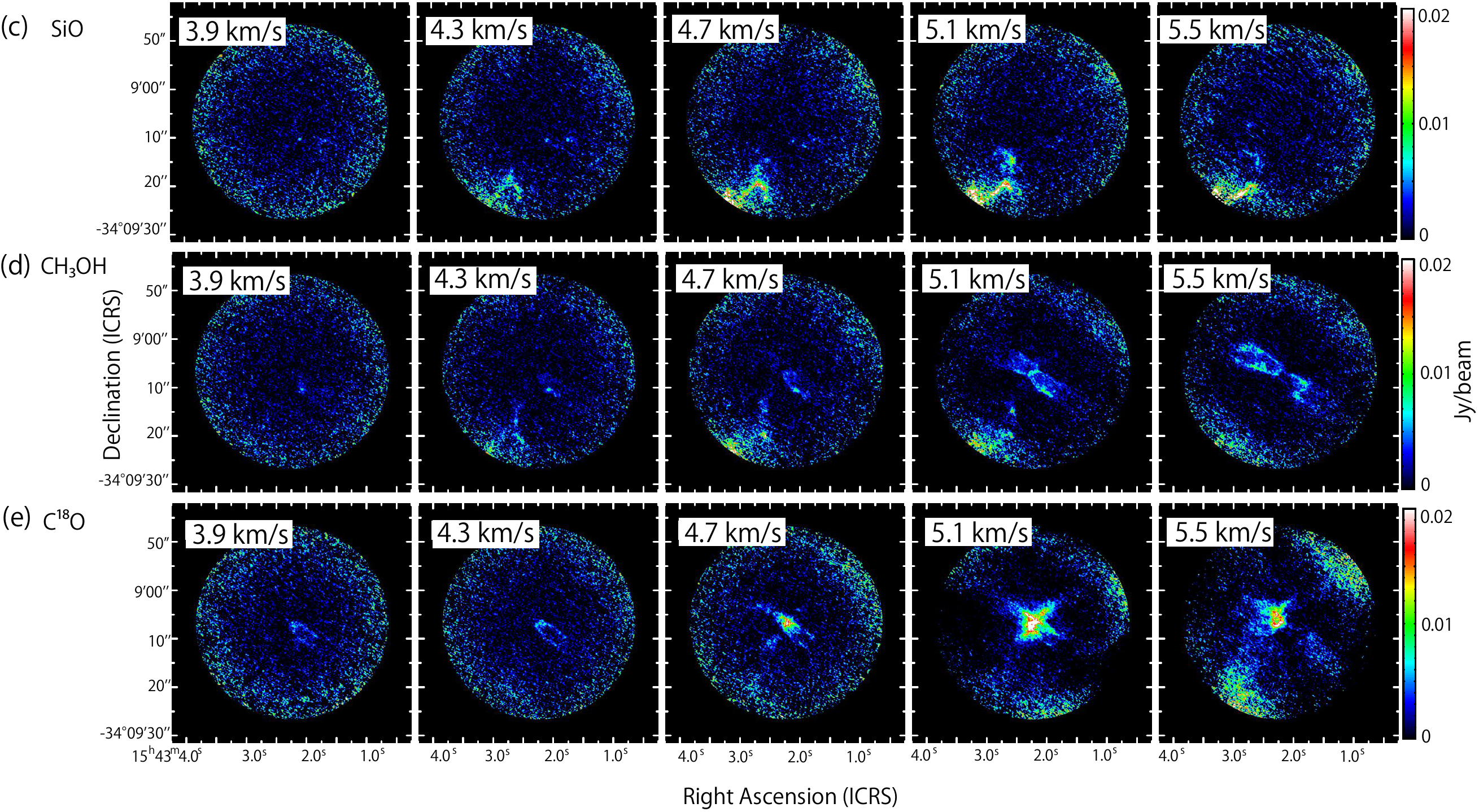}
\caption{Continued.\label{channel_1}}
\end{minipage}}
\end{figure}

\begin{figure}[h!]
\centering
\includegraphics[scale=0.5]{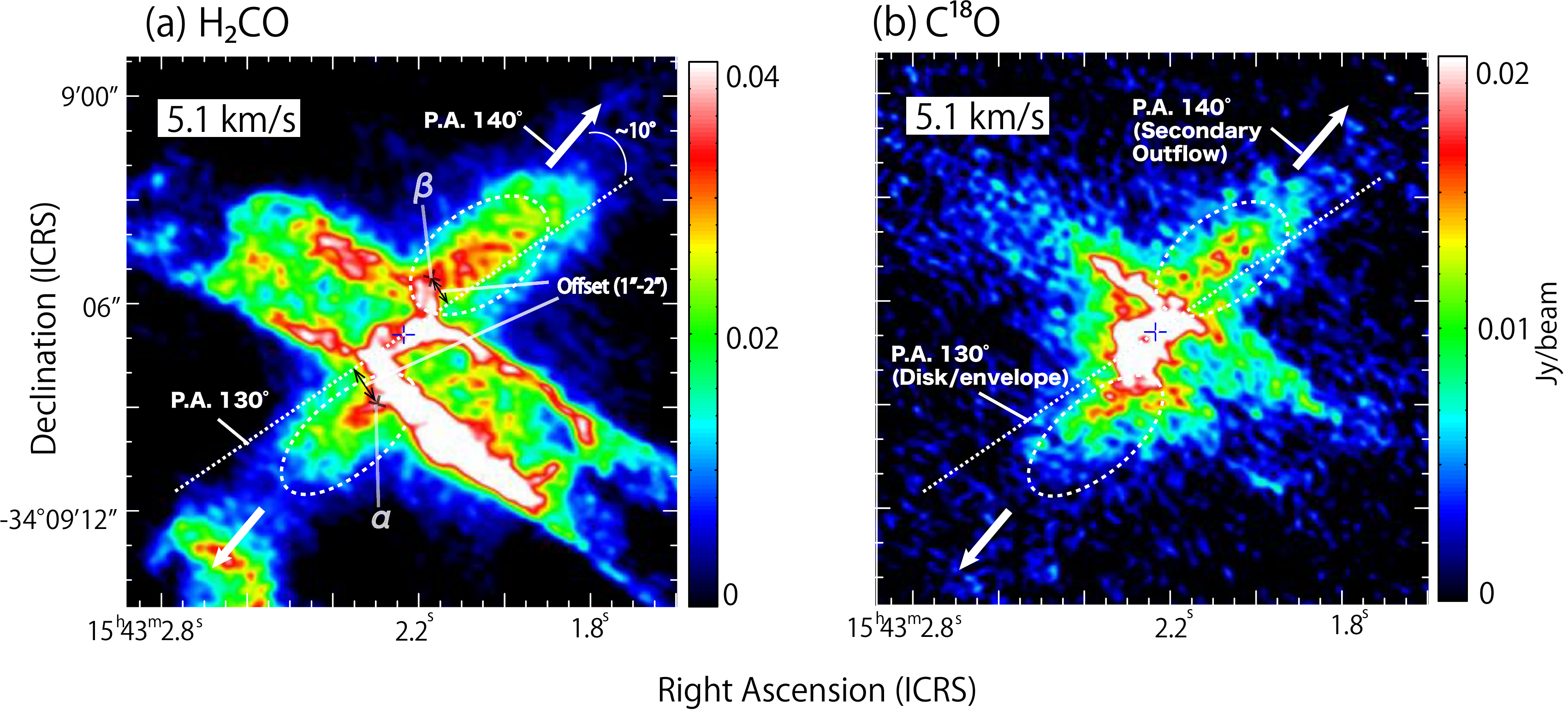}
\caption{Blow-ups of the 5.1 \kms\ channel maps from Figures \ref{channel_1}b and \ref{channel_1}e. The white dashed ovals indicate the linear structure. The arrows indicate the direction of the secondary outflow (P.A. 140\degr). The dashed lines represent the disk/envelope axis (P.A. 130\degr). $\alpha$ and $\beta$ represent the origin points of the linear structure described in Section 3.4. They represent the intersection points between the linear structure and the primary outflow cavity. The blue cross marks indicate the continuum peak position. \label{shift}}
\end{figure}

\begin{figure}[h!]
\centering
\includegraphics[scale=0.5]{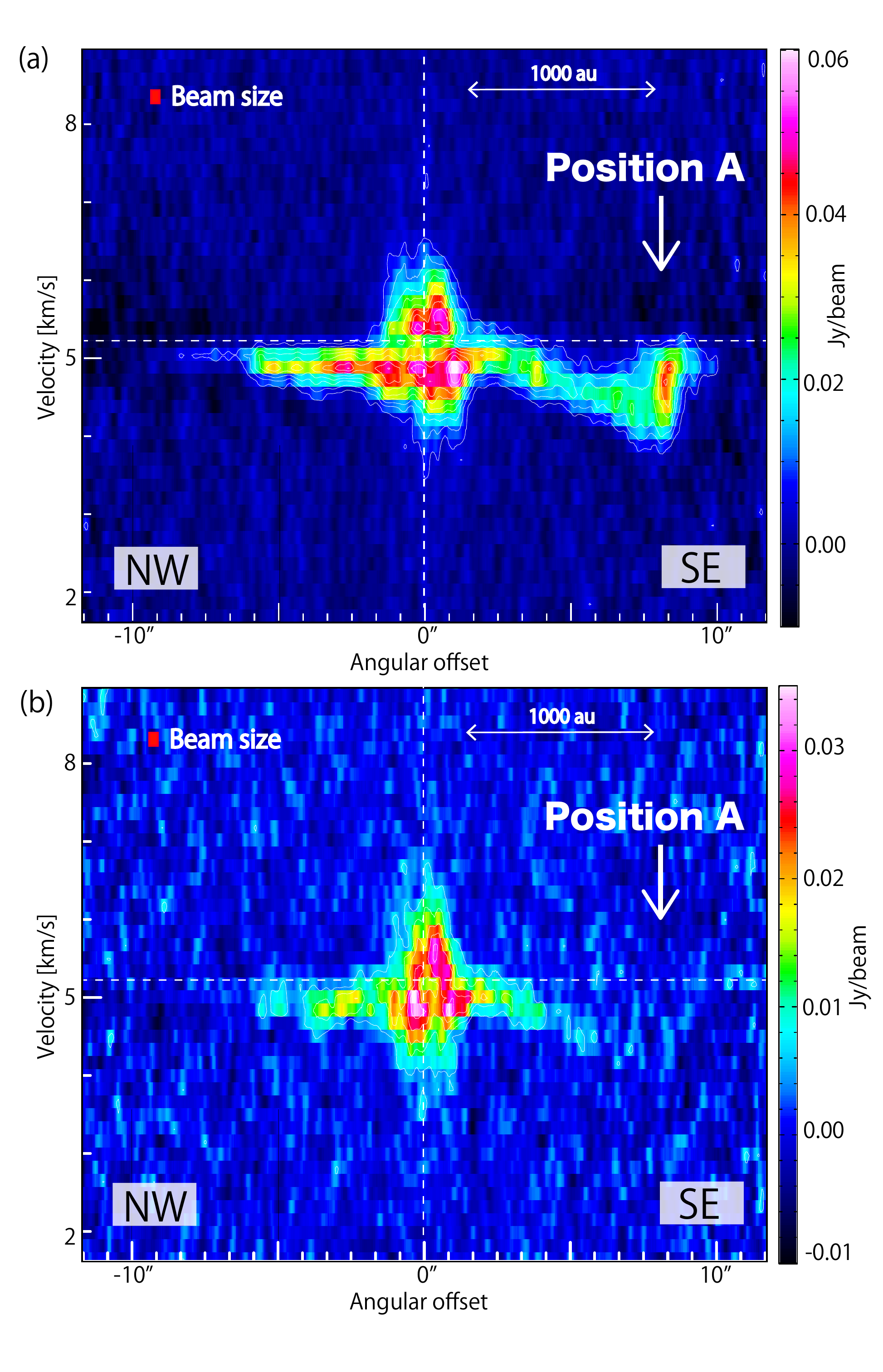}
\caption{
(a,b) PV diagrams for \hhco\ and \co\ along the northwest to the southeast axis (P.A. 140\degr; Figure \ref{moment_1}a). The origin shows the continuum peak position. The horizontal dashed line indicates the systemic velocity of the protostellar core (5.2 \kms). The cut width is 1\farcs0.
 \label{pv}}
\end{figure}

\end{document}